\documentclass[aps,prl,showpacs,twocolumn,floats,superscriptaddress]
{revtex4}
\usepackage{graphicx}
\usepackage{bm}
\usepackage{color}
\makeatletter
\def\lsim{\mathrel{\mathpalette\gl@align<}}
\def\gsim{\mathrel{\mathpalette\gl@align>}}
\def\gl@align#1#2{\lower.6ex\vbox
{\baselineskip\z@skip\lineskip\z@
\ialign{$\m@th#1\hfil##\hfil$\crcr#2\crcr\sim\crcr}}}

\newcommand{\rr}{\ensuremath{{\bf r}}}
\newcommand{\rp}{\ensuremath{{\bf r'}}}
\newcommand{\nbr}{\ensuremath{\langle {\bf r} {\bf r'}\rangle}}
\newcommand{\nbrr} {\ensuremath{\langle {\bf r}_1 {\bf r}_2\rangle}}
\newcommand{\abar}{\ensuremath{\bar{\alpha}}}

\makeatother

\begin{document}

\title{Role of trap-induced scales in non-equilibrium dynamics of strongly interacting trapped bosons}

\author{Anirban Dutta}
\affiliation{Theoretical Physics Department, Indian Association for
the Cultivation of Science, Jadavpur, Kolkata-700032, India.}

\author{Rajdeep Sensarma}
\affiliation{Department of Theoretical Physics, Tata Institute of
Fundamental Research, Mumbai-400005, India.}

\author{K. Sengupta}
\affiliation{Theoretical Physics Department, Indian Association for
the Cultivation of Science, Jadavpur, Kolkata-700032, India.}

\date{\today}

\begin{abstract}

We use a time-dependent hopping expansion technique to study the
non-equilibrium dynamics of strongly interacting bosons in an
optical lattice in the presence of a harmonic trap characterized by
a force constant $K$. We show that after a sudden quench of the
hopping amplitude $J$ across the superfluid (SF)-Mott insulator(MI)
transition, the SF order parameter $|\Delta_{\bf r}(t)|$ and the
local density fluctuation $\delta n_{\bf r}(t)$ exhibit sudden
decoherence beyond a trap-induced time scale $T_0 \sim K^{-1/2}$. We
also show that after a slow linear ramp down of $J$, $|\Delta_{\bf
r}|$ and the boson defect density $P_{\bf r}$ display a novel
non-monotonic spatial profile. Both these phenomena can be explained
as consequences of trap-induced time and length scales affecting the
dynamics and can be tested by concrete experiments.

\end{abstract}


\maketitle

The study of non-equilibrium dynamics of closed quantum many-body
systems has gained tremendous momentum in recent years \cite{pol1},
mainly due to experiments on ultracold atoms in optical lattices
\cite{bloch1,kinoshita1,bakr1,Esslinger,cn1}. Isolation from
external baths and long timescales for dynamics in these ultra low
temperature systems make it easy to follow the system in real time
without ultrafast probes. Ultracold atoms can be used to emulate
several strongly correlated quantum Hamiltonians like the Ising and
the Bose-Hubbard (BH) models, which are known to undergo quantum
phase transitions \cite{bloch2,sachdev1, sachdev2} as a function of
easily tuneable parameters. Experiments on these systems provide a
unique opportunity to study non-equilibrium dynamics of strongly
interacting quantum many-body systems in the vicinity of quantum
critical points.

The experimental setup of ultracold atoms inevitably has a harmonic
trapping potential, which provides the largest lengthscale in the
problem. In equilibrium, the trap: (a) smears out phase transitions
into crossovers, as the long wavelength low energy ``critical''
fluctuations are cut-off at the trap lengthscale and (b) realizes
multiple phases like a Mott insulator (MI) and a superfluid (SF)
coexisting in different regions of the trap, leading to interesting
phase boundaries between them. The dynamics of these strongly
interacting systems in the presence of confining potentials is an
interesting topic, which has implications over a wide range of
physical phenomenon from dynamics of hot dense QCD matter to nuclear
reactions to dynamics of early universe and so on. While the
equilibrium phase diagram of ultracold atoms in presence of
confining potentials is well studied, the interplay of strong
interactions and confinement in the dynamics of these systems is
open to new descriptions.

The BH model, emulated by the ultracold atom systems, consists of
bosons hopping on a lattice with a hopping amplitude $J$ and
interacting with a local repulsion $U$. It supports a
superfluid-insulator (SI) quantum phase transition as a function of
the parameter $J/U$, with the quantum critical point occurring in
the strong coupling regime $U \gg J$. In this Letter, we study the
non-equilibrium dynamics of this system in the presence of a
harmonic trap characterized by a force constant $K$, when we change
the hopping parameter $J$ from an initial high value (in the
equilibrium superfluid phase) to a low value (in the equilibrium
Mott insulator phase) across the quantum phase transition.

We first look at the evolution of the system after an instantaneous
quench of the hopping parameter. Initially the local superfluid
order parameter $|\Delta_{r}|$ and the local density fluctuations
$\delta n_r$ follow the well known collapse and revival dynamics of
the homogeneous system up to a trap induced time scale $T_0 \sim
K^{-1/2}$. Beyond this timescale, the oscillations decohere suddenly
and the system settles into a steady state. We also look at the
evolution of the system through a linear ramp-down of the hopping
parameter $J(t)$ with a rate $\tau^{-1}$. We find that at the end of
slow ramps ($\tau U \gg 1$), both $|\Delta_{r}|$ and the local
defect density $P_r$ display novel non-monotonic spatial profile as
a function of $r$, which characterizes the highly non-equilibrium
state produced at the end of the ramp. We provide a semi-analytic
explanation for both the $K$ dependence of $T_0$ and the
non-monotonic spatial profile of $|\Delta_r|$ and $P_r$ after the
ramp  in terms of trap-induced length and time scales which affect
the dynamics of these systems in presence of confining potential,
and suggest concrete experiments to test our theory.

We would like to note that, while several numerical and analytical
methods are used to study the equilibrium properties of the SI
transition in BH model for $d>1$ \cite{mft1,stc1,stc2,qmc}, none of
these can describe its non-equilibrium dynamics beyond the
mean-field level. Recently, some progress has been made in this
direction in Ref.\ \cite{stc2}, but treatment of dynamics of
non-uniform systems are beyond that method. To the best of our
knowledge, methods to address non-equilibrium dynamics of
inhomogeneous BH model beyond mean-field theory do not exist for
$d>1$ \cite{collath1,natu1}. Here, we develop a hopping expansion
technique for strongly coupled bosons in an optical lattice in the
presence of a spatially varying potential which treats the
equilibrium and non-equilibrium properties of the system on an equal
footing.

The BH model in the presence of a trap is given by
\begin{equation}
H = -J \sum_{\langle {\bf r r'}\rangle} (b_{\bf r}^{\dagger} b_{\bf
r'} +h.c).+ \sum_{\bf r} \frac{U}{2} {\hat n}_{\bf r} ( {\hat
n}_{\bf r} -1)- \mu_{\bf r} {\hat n}_{\bf r}, \label{eqham1}
\end{equation}
where $b_{\bf r}$ annihilates a boson at lattice site ${\bf r}$ and
${\hat n}_{\bf r}=b_{\bf r}^{\dagger} b_{\bf r}$. Here, $\mu_{\bf r}
= \mu_0 - K |{\bf r}|^2/2$ is the effective chemical potential at
site  ${\bf r}$, $K$ is the force constant of the harmonic trap
potential, and the central value $\mu_0$ controls the total density.
For concreteness, we will consider nearest neighbor hopping of
bosons on a square lattice.

The BH hamiltonian can be divided into a local part $H_0$,
consisting of the Hubbard repulsion and the effective chemical
potential term, and the kinetic hopping term $T$. In the strong
coupling regime, a perturbation expansion in the hopping terms is
obtained in the following way: Consider neighboring sites ${\bf r}$
and  ${\bf r'}$ with occupation numbers $n_{\bf r}$ and $n_{\bf r'}=
n_{\bf r}-\alpha +1$, where $\alpha$ is an integer. When a boson
hops from ${\bf r'}$ to ${\bf r}$, the energy change $\Delta E_{\rr
\rp}^{\alpha} = \alpha U - \mu_{\bf r}+\mu_{\bf r'}$. It is thus
useful to consider the hopping terms corresponding to the same
$\alpha$ together and write: $T = \sum_{\nbr} T_{\rr\rp} =
\sum_{\nbr \alpha} (T_{\rr\rp}^{\alpha} + T_{\rp\rr}^{-\alpha})$,
where
\begin{eqnarray}
T_{\rr\rp}^{\alpha} &=& -J \sum_{n_{\bf r}}\sqrt{(n_{\bf
r}+1)(n_{\bf r} +1- \alpha)} |n_{\bf r}+1, n_{\bf r} - \alpha
\rangle \nonumber\\
&& \times \langle n_{\bf r},n_{\bf r} - \alpha +1|.  \label{hopp1}
\end{eqnarray}

For a given pair of $\rr$ and $\rp$, the low energy process
corresponds to $\abar_{\rr\rp}$, for which $\Delta
E^{\abar_{\rr\rp}}_{\rr\rp} < \gamma J$, where $\gamma$ is a number
${\cal O}(1)$~\cite{footnotegamma}. The rest of the terms are taken
as high energy processes to be eliminated by a canonical
transformation. Note that there can be pairs of sites where all
hopping processes are high energy terms, and hence there are no
terms ${\cal O}(J)$ in the effective Hamiltonian between these
sites. A binary variable $\eta_{\rr\rp}$, which is 1(0) if the
corresponding bond has (does not have) a low energy hopping term, is
used to keep track of this.

Having identified the high-energy hopping processes for which
$\alpha_{\rr\rp} \ne {\abar }_{\rr\rp}$ , one can now design a
canonical transformation operator $S$ which eliminate these
processes perturbatively to obtain an effective Hamiltonian $H_{\rm
eff} = \exp (iS) H \exp(-iS)$. To linear order in $J$, this requires
$[H_0,iS]= \sum_{\nbr \alpha \ne {\abar
}_{\rr\rp}}(T_{\rr\rp}^{\alpha} + T_{\rp\rr}^{-\alpha})$ and yields
\begin{eqnarray}
iS &=& \sum_{\nbr} \sum_{\alpha \ne {\abar }_{\rr\rp}}
(T_{\rr\rp}^{\alpha}-T_{\rp\rr}^{-\alpha})/\Delta
E_{\rr\rp}^{\alpha}. \label{candef}
\end{eqnarray}
where we have used the fact that $\abar_{\rr\rp}=-\abar_{\rp\rr}$.

\begin{figure}[t!]
\begin{center}
\includegraphics[width=0.5\columnwidth]{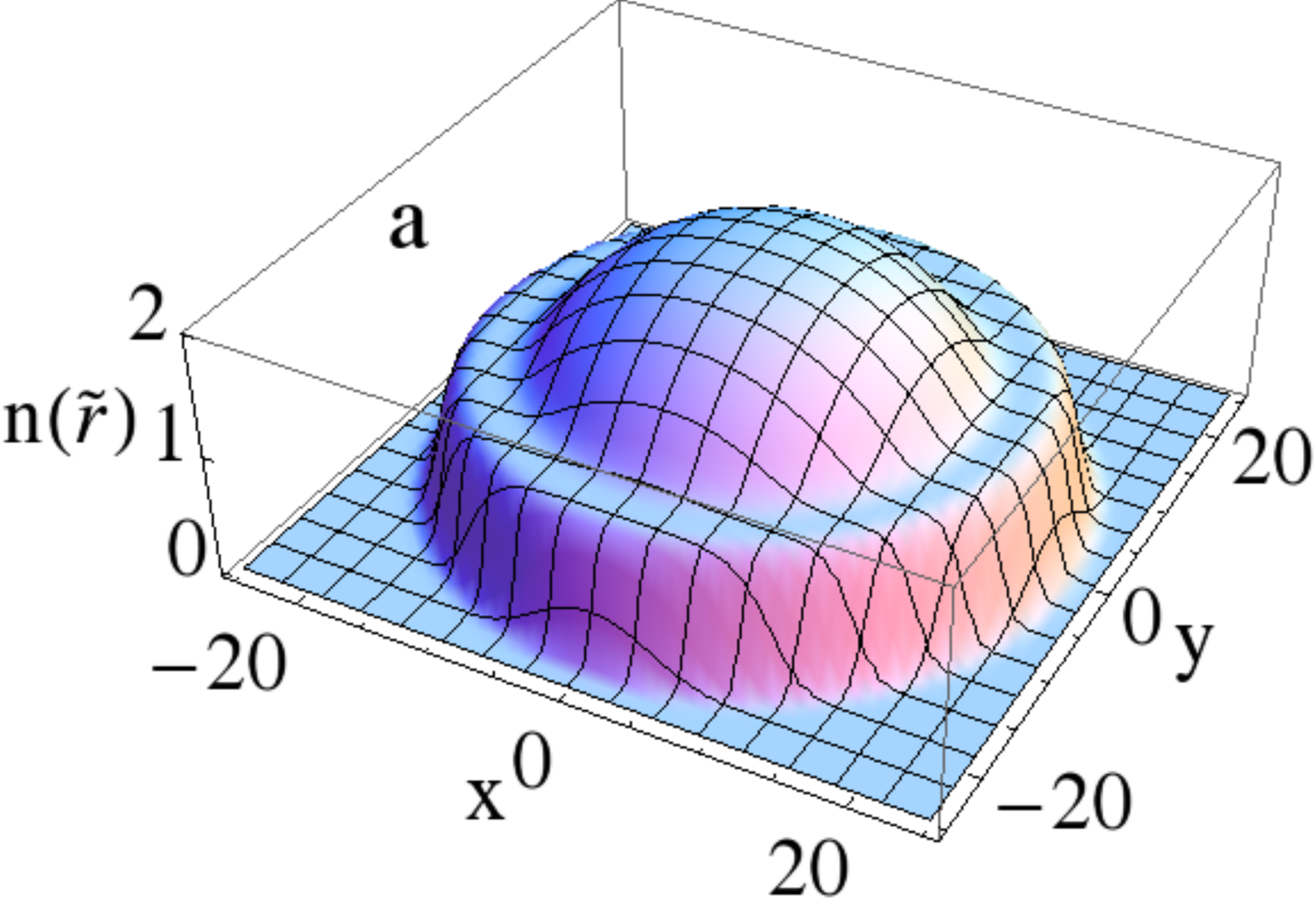}
\includegraphics[width=0.45\columnwidth]{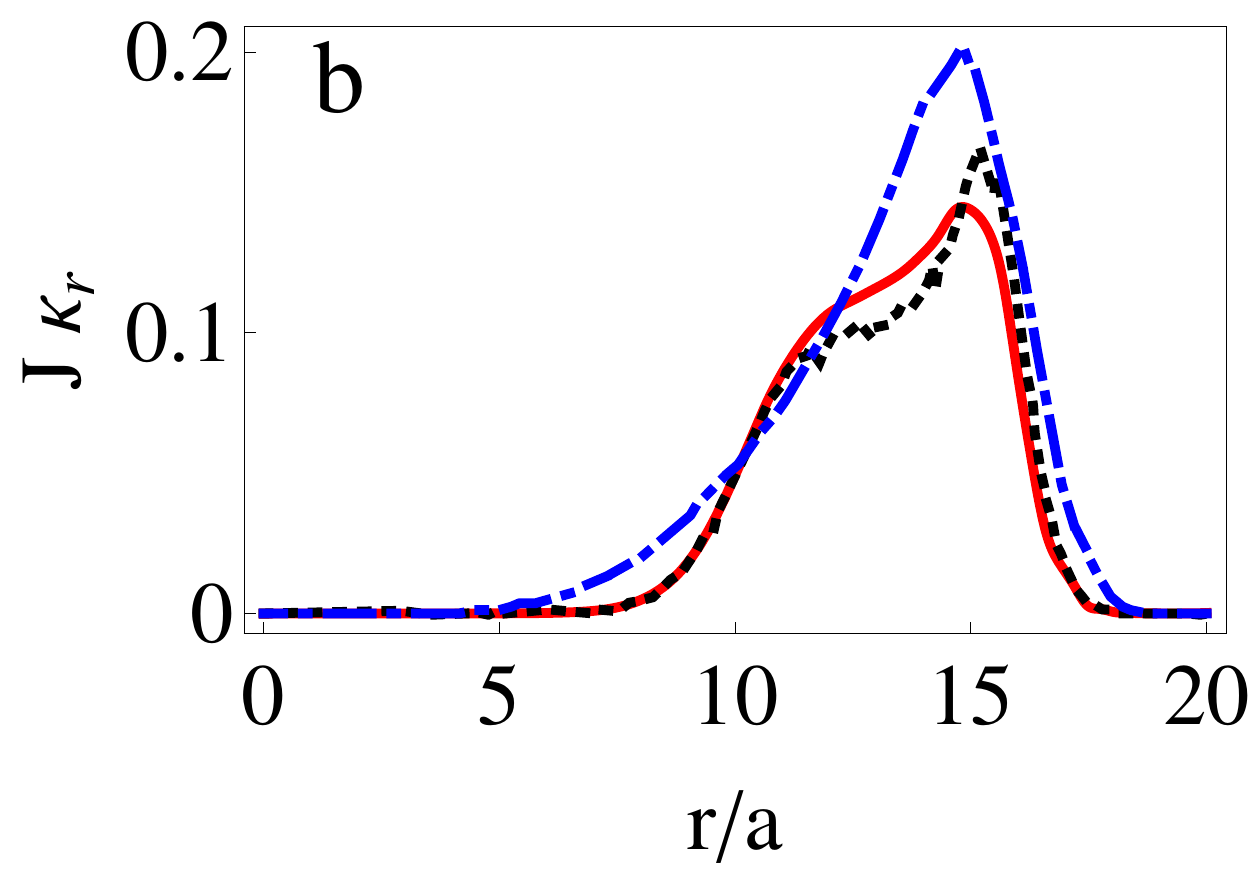}
\end{center}
\caption{(Color online) (a): Density profile of the bosons in a 2D
harmonic trap with $\mu_0=1.4U$ and $Ka^2=0.006 U$ (b): Plot of $J
\kappa_{\bf r}$ (red solid curve) as a function of the distance from
the trap center, $r_0$, for $\mu_0/U=0.37$ and $K a^2=0.002U$
showing comparison with QMC data (black dotted curve) and mean-field
theory (blue dash-dotted curve). $J=0.04 U$ and $N= 51 \times 51$
for both plots.}\label{fig1}
\end{figure}

Using the above $S$, the effective low-energy Hamiltonian $H_{\rm
eff}$ for the trapped bosons, to ${\cal O}(z^2J^2/U^2)$, is
\begin{eqnarray}
&& H_{\rm eff} = H_0 + \sum_{\nbr}
[T_{\rr\rp}^{{\abar}_{\rr\rp}}+T_{\rp\rr}^{{-\abar}_{\rr\rp}}]
\eta_{\rr\rp} \\
&& +\displaystyle \sum_{\nbr \nbrr}\eta_{\rr\rp} \sum_{\alpha \ne
{\abar }_{\rr_1\rr_2}} \frac{ \left[
T_{\rr_1\rr_2}^{\alpha}-T_{\rr_2\rr_1}^{-\alpha},
T_{\rr\rp}^{\abar_{\rr\rp}}+T_{\rp\rr}^{-\abar_{\rp\rr}}\right]}{\Delta
E_{\rr_1\rr_2}^{\alpha} }
\nonumber\\
&& +  \displaystyle \sum_{\nbr \nbrr}\sum_{\alpha \ne {\abar
}_{\rr_1\rr_2}
  \beta \ne {\bar{\beta} }_{\rr\rp}}
\frac{ \left[ T_{\rr_1\rr_2}^{\alpha}-T_{\rr_2\rr_1}^{-\alpha},
T_{\rr\rp}^{\bar{\beta}}+T_{\rp\rr}^{-\bar{\beta}}\right]}{2\Delta
E_{\rr_1\rr_2}^{\alpha} }. \nonumber\
\label{effham}
\end{eqnarray}
We first use this formalism to look at the equilibrium ground state
of the trapped system. We use a variational wavefunction ansatz,
$|\psi\rangle = \exp(-i S) |\psi'\rangle$, where $|\psi'\rangle =
\prod_{\bf r} \sum_{n_{\bf r}} f_{n_{\bf r}}^{\bf r} |n_{\bf
r}\rangle$ is a Gutzwiller wavefunction, $n_{\bf r}$ is the local
boson occupation number basis, and the variational Gutzwiller
coefficients $f_{n_{\bf r}}^{\bf r}$ are determined by minimizing
the ground state energy $E_G= \langle \psi|H|\psi\rangle = \langle
\psi'|H_{\rm eff} |\psi'\rangle + {\cal O}(z^3J^3/U^2)$
\cite{supp1}. We note that in our approach, $|\psi\rangle$, unlike
$|\psi'\rangle$, retains spatial correlations due to the $\exp(iS)$
factor. The ground state expectation value of any operator ${\hat
O}$ to ${\cal O}(J^2/U^2)$ is then given by
\begin{eqnarray}
\langle \psi |{\hat O}|\psi\rangle &=& \langle \psi'|e^{iS} {\hat O}
e^{-iS}|\psi'\rangle + {\cal O}(z^3J^3/U^3). \label{expval}
\end{eqnarray}

The ground state density profile for a system of $51 \times 51$
lattice sites is shown in Fig.\ \ref{fig1}(a). It shows the wedding
cake structure with the $n_{\bf r}=\langle \hat{n}_{\bf r}\rangle=2$
central Mott lobe surrounded by a SF ring and then the $ n_{\bf
r}=1$ Mott lobe, as we go towards the edge of the trap. To make
quantitative comparison between this method, Quantum Monte Carlo
(QMC) \cite{wessel1} and mean-field theory\cite{mft1},
 we plot in Fig.~\ref{fig1}(b), the local compressibility $\kappa_{\bf
r} = \langle \psi |{\hat n}_{\bf r}^2 |\psi\rangle - n_{\bf r}^2$ of
the system  as a function of $r_0=|{\bf r}|$. In this case, we use
$\mu_0=0.37 U$ so that we only have a $n=1$ central Mott lobe. The
comparison shows that our method provides a more accurate match with
QMC data than the mean-field theory in the MI and MI-SF transition
regions.

We now turn to the description of the non-equilibrium dynamics of
this system as the hopping is changed according to an arbitrary
protocol $J(t)$. The Schrodinger equation is given by $i \hbar
\partial_t |\psi (t) \rangle = H[J(t)] |\psi(t)\rangle$. Following
Ref.~\cite{stc2}, we work with a time-dependent canonical
transformation $S[J(t)]$, i.e.
$|\psi(t)\rangle=e^{iS[J(t)]}|\psi^{'}(t)\rangle$. The basic idea is
to keep the fast oscillating (high energy) terms within the
canonical transformation, so that $|\psi^{'}(t)\rangle$ can encode
the slow motion generated by the time-dependent effective
Hamiltonian $H_{\rm eff}[J(t)]$ (given by Eq.\ \ref{effham} with
$J\to J(t)$).  The Schrodinger equation then reduces to
\begin{eqnarray}
(i \hbar \partial_t + \hbar
\partial S/\partial t ) |\psi'(t)\rangle = H_{\rm eff}[J(t)]
|\psi'(t)\rangle.
\end{eqnarray}
We then use a time dependent Gutzwiller ansatz $|\psi'(t)\rangle =
\prod_{\bf r} \sum_{n_{\bf r}} f_{n_{\bf r}}^{\bf r}(t) |n_{\bf
r}\rangle$ and obtain the differential equations for $f_{n_{\bf
r}}^{\bf r}(t)$ \cite{supp1}. These are solved numerically to obtain
the time dependent state $|\psi(t)\rangle$ \cite{commentzz}.

\begin{figure}[t!]
\begin{center}
\includegraphics[width=0.49\columnwidth]{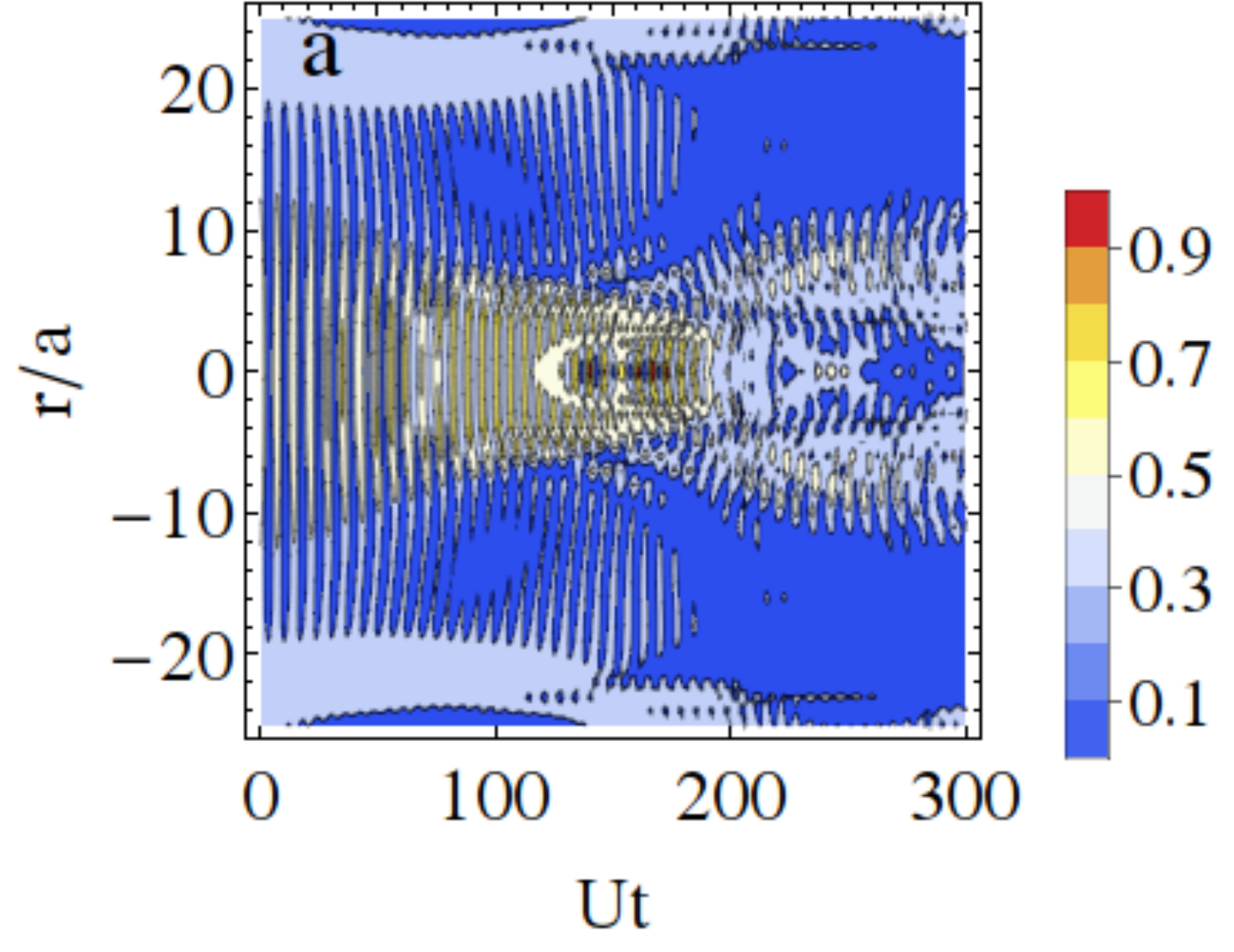}
\includegraphics[width=0.49\columnwidth]{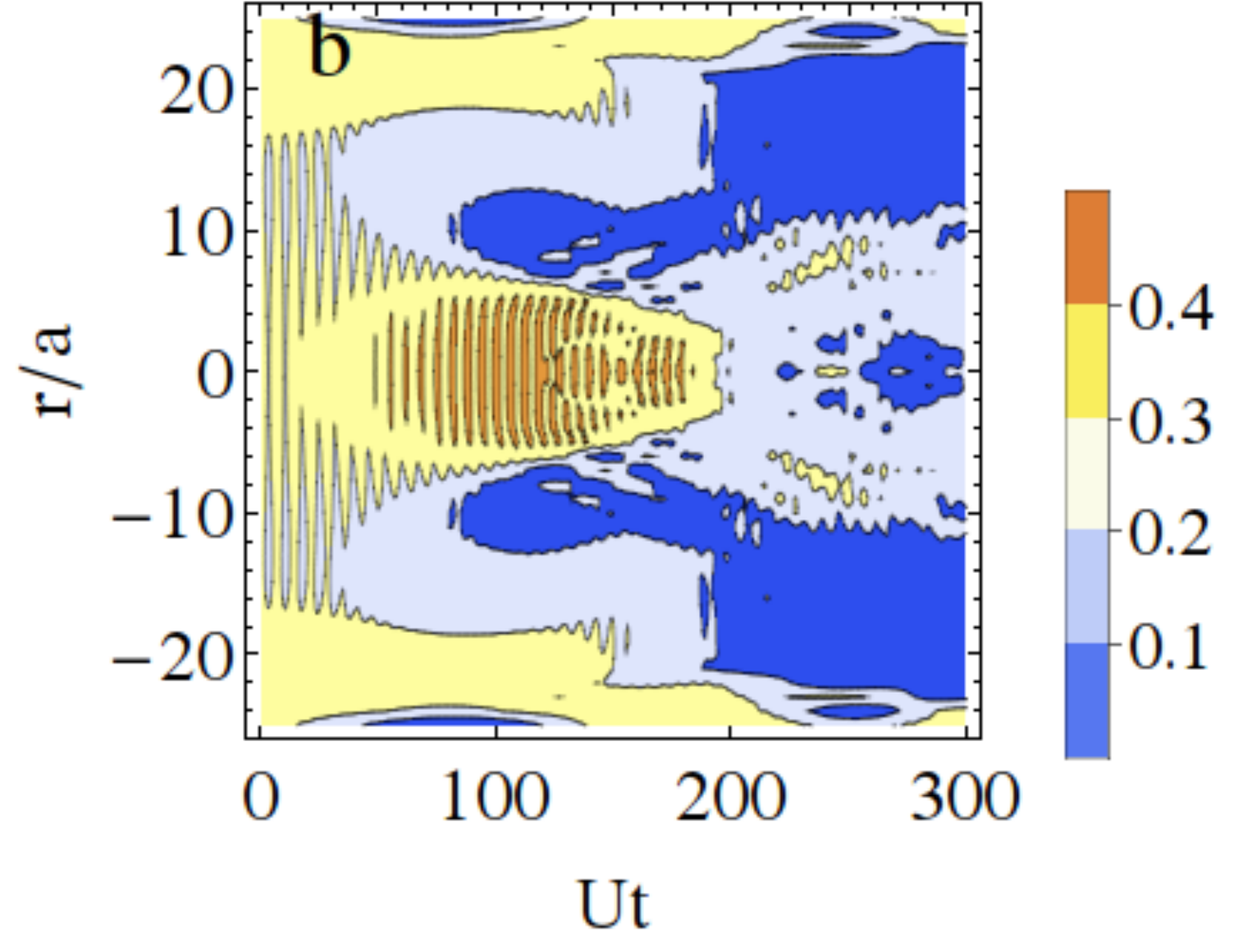}
\includegraphics[width=0.45\columnwidth]{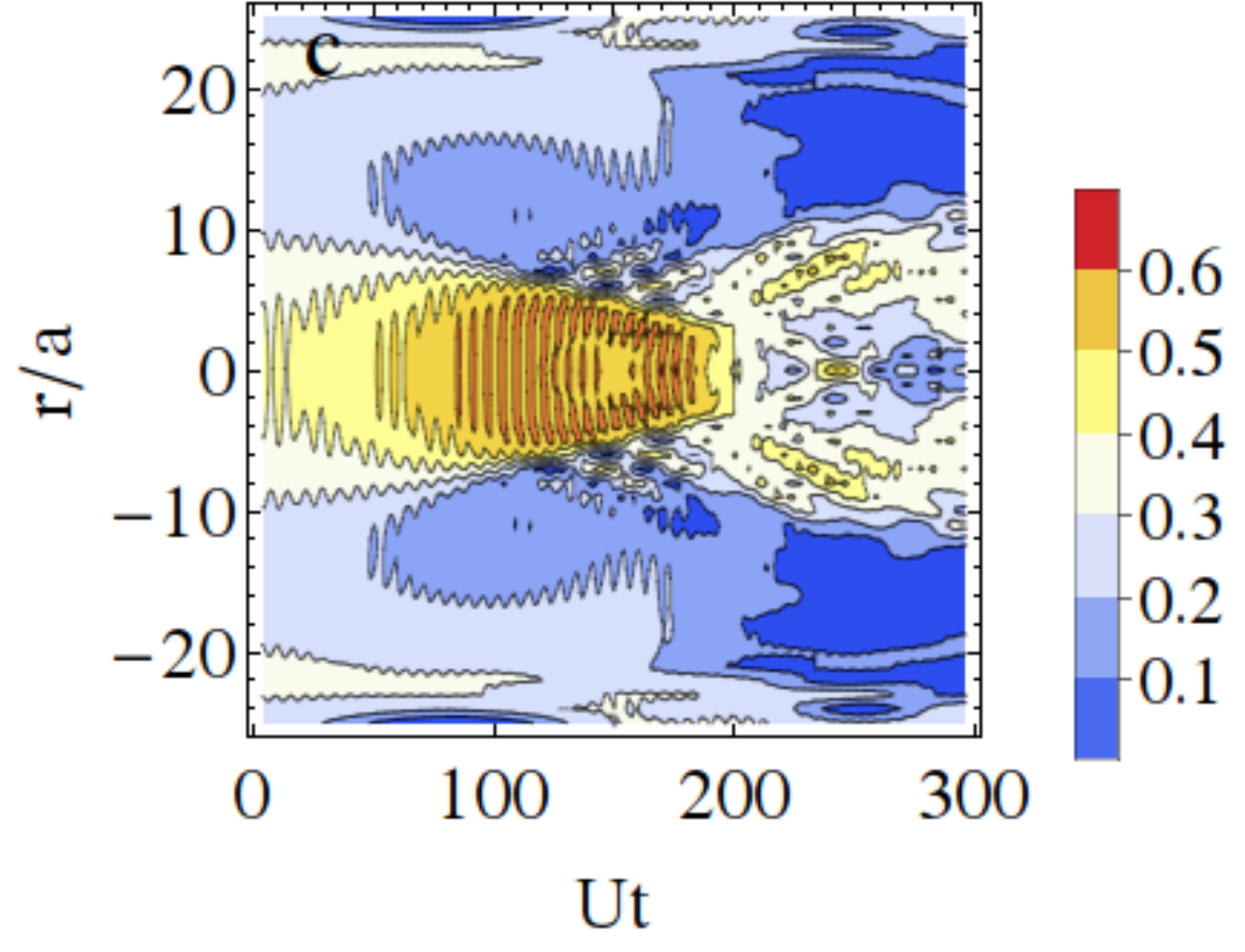}
\includegraphics[width=0.45\columnwidth]{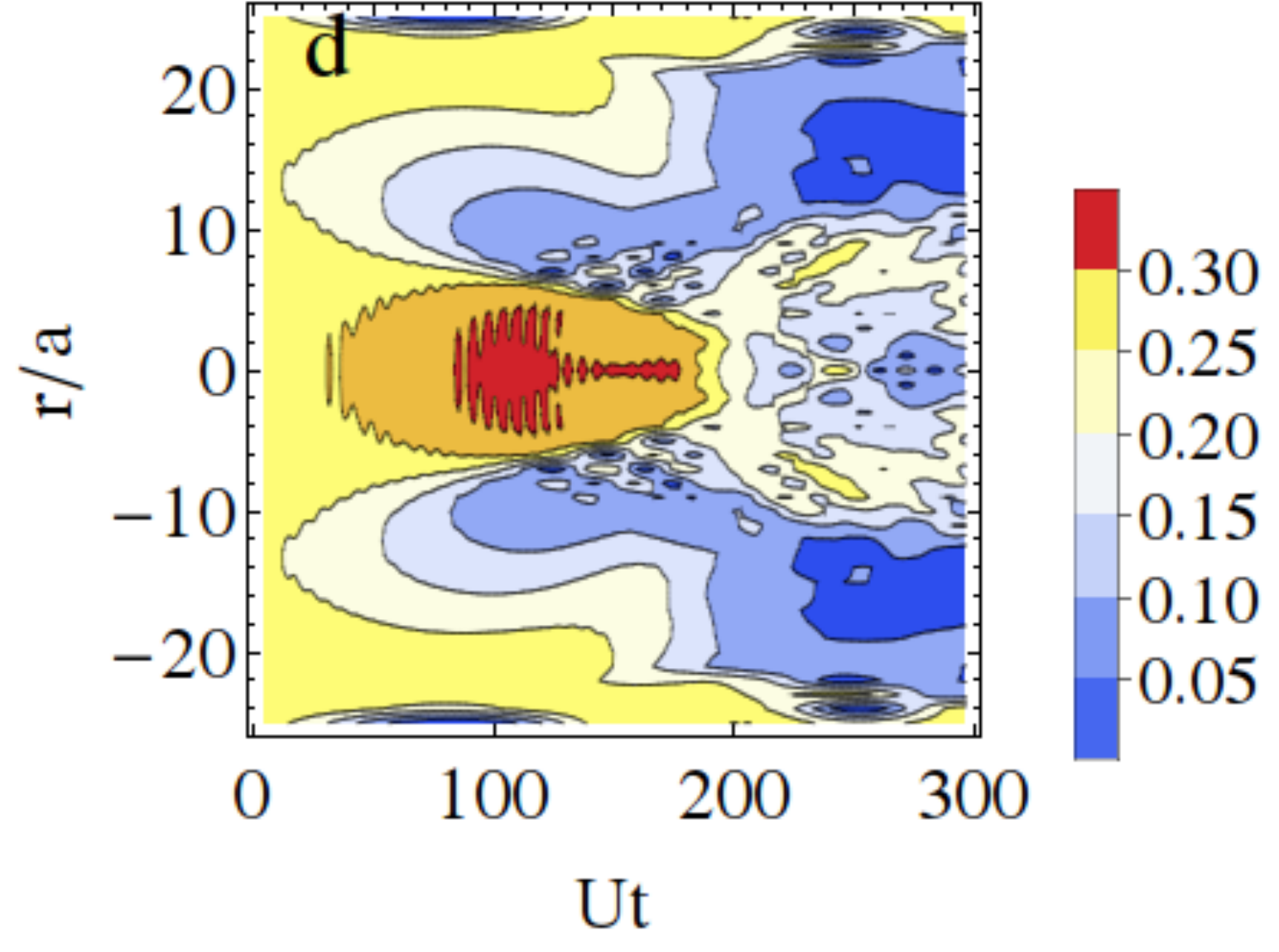}
\end{center}
\caption{(Color online) Time evolution after a sudden quench of $J$
from $J_i=0.1 U$ to $J_f=0.02 U$: (a) $|\Delta_{r}(t)|$ and (b)
$\delta n_{r}(t)$ as a function of $r$ and $t$. (c) and (d): Same as
(a) and (b) respectively, but integrated over a timescale $\delta
t=2\pi/U$ to show the slow dynamics clearly.} \label{fig2}
\end{figure}

First, we concentrate on the sudden quench of $J$ from $J=J_i$,
where the bosons in the trap center are in the SF phase, to $J=J_f$,
where the ground state at the trap center is a MI with ${\bar n}=1$.
In Fig.\ \ref{fig2}(a), we plot the spatio-temporal profile of
$|\Delta_r(t)|$, where $\Delta_r(t)=\langle
\psi(t)|b_r|\psi(t)\rangle$ is the local superfluid order parameter,
and $r$ is the distance from the trap center along $(1,0)$. It shows
prominent oscillations with a frequency $\sim U^{-1}$, corresponding
to the coherent collapse and revival of the superfluid state in the
center. Around $T_0 \sim 1/K^{1/2}$, these oscillations suddenly
decohere very fast and soon the system settles into a steady state
pattern. This decoherence is not the exponential decay due to
hopping, but is precipitated by a catastrophic event which
immediately causes loss of coherence. A similar pattern is seen in
the local density fluctuations $\delta n_r(t)$, plotted in Fig.\
\ref{fig2}(b). In Figs.\ \ref{fig2}(c) and (d), we plot the
spatio-temporal profile of $|\Delta_r(t)|$ and $\delta n_r(t)$,
after smearing over a time grid $\delta t \sim U^{-1}$ to clearly
visualize long time-scale dynamics of the system, showing how it
settles into the steady state.
\begin{figure}[t!]
\begin{center}
\includegraphics[width=0.49\columnwidth]{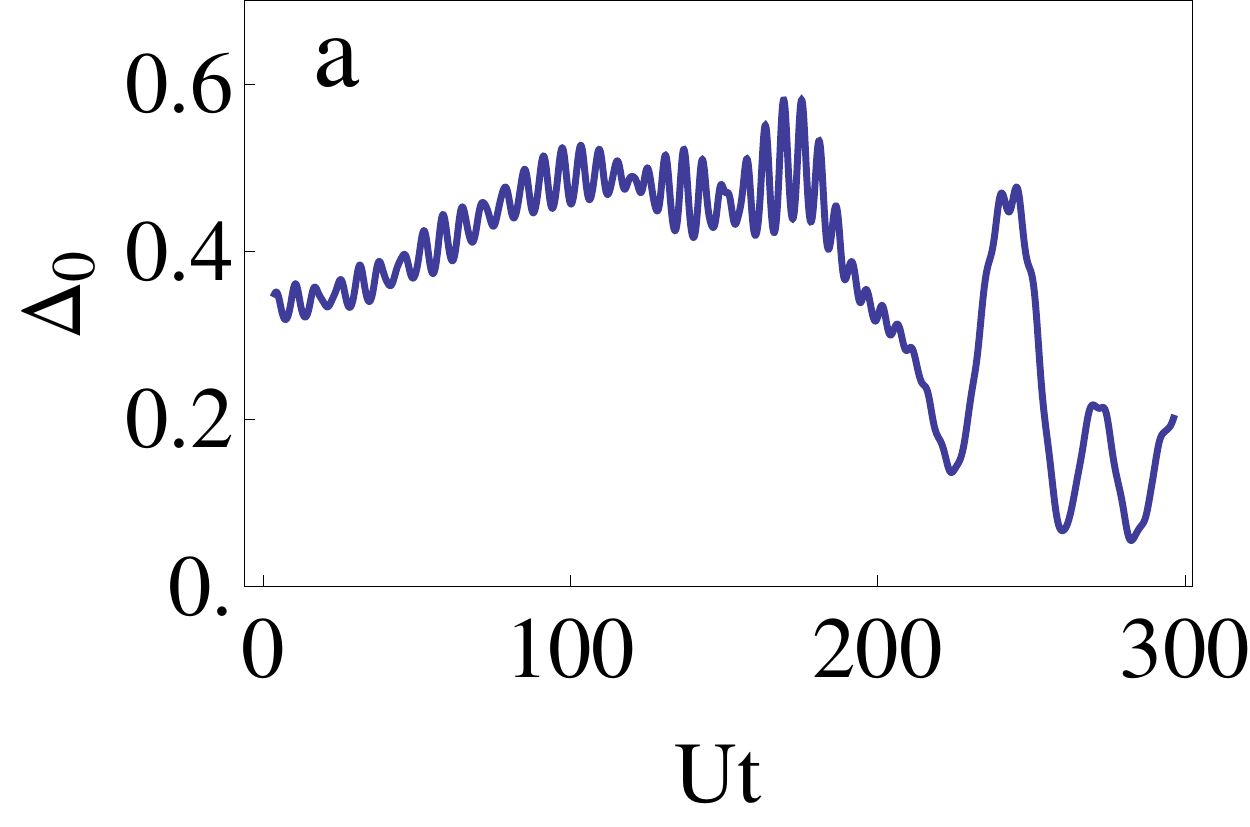}
\includegraphics[width=0.49\columnwidth]{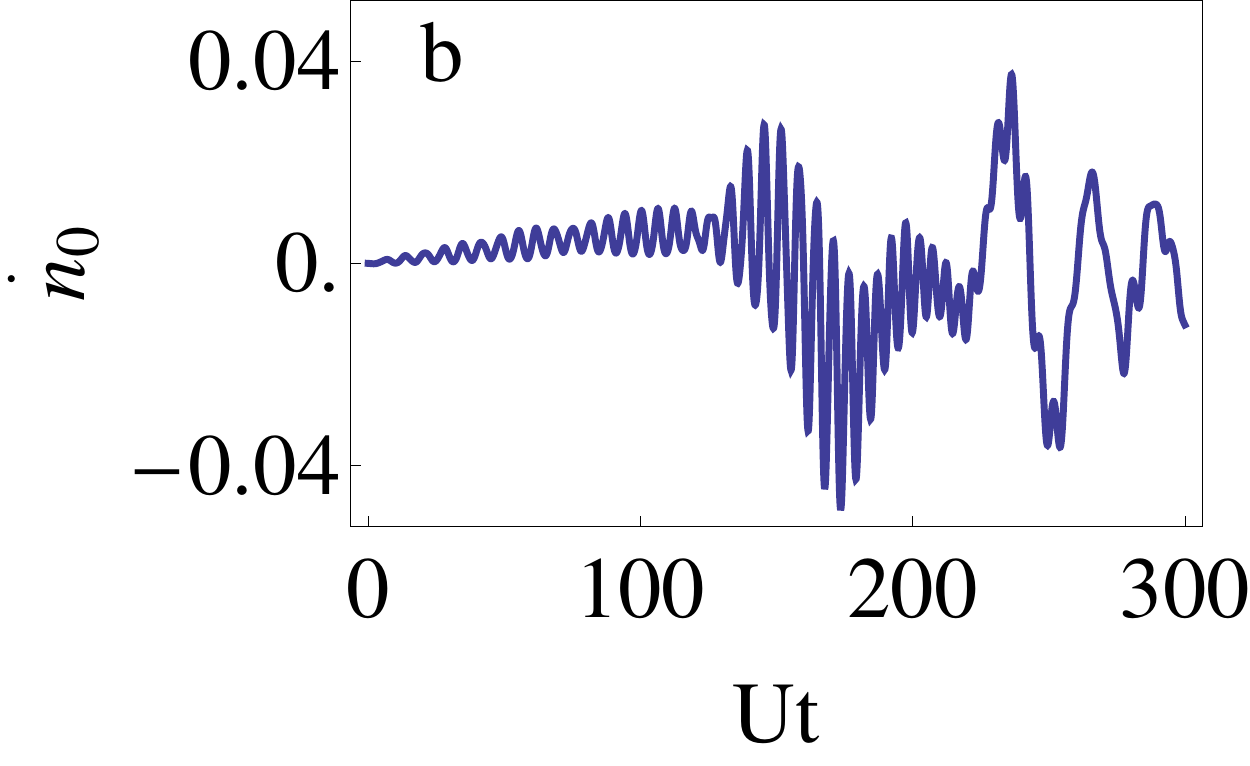}
\includegraphics[width=0.45\columnwidth]{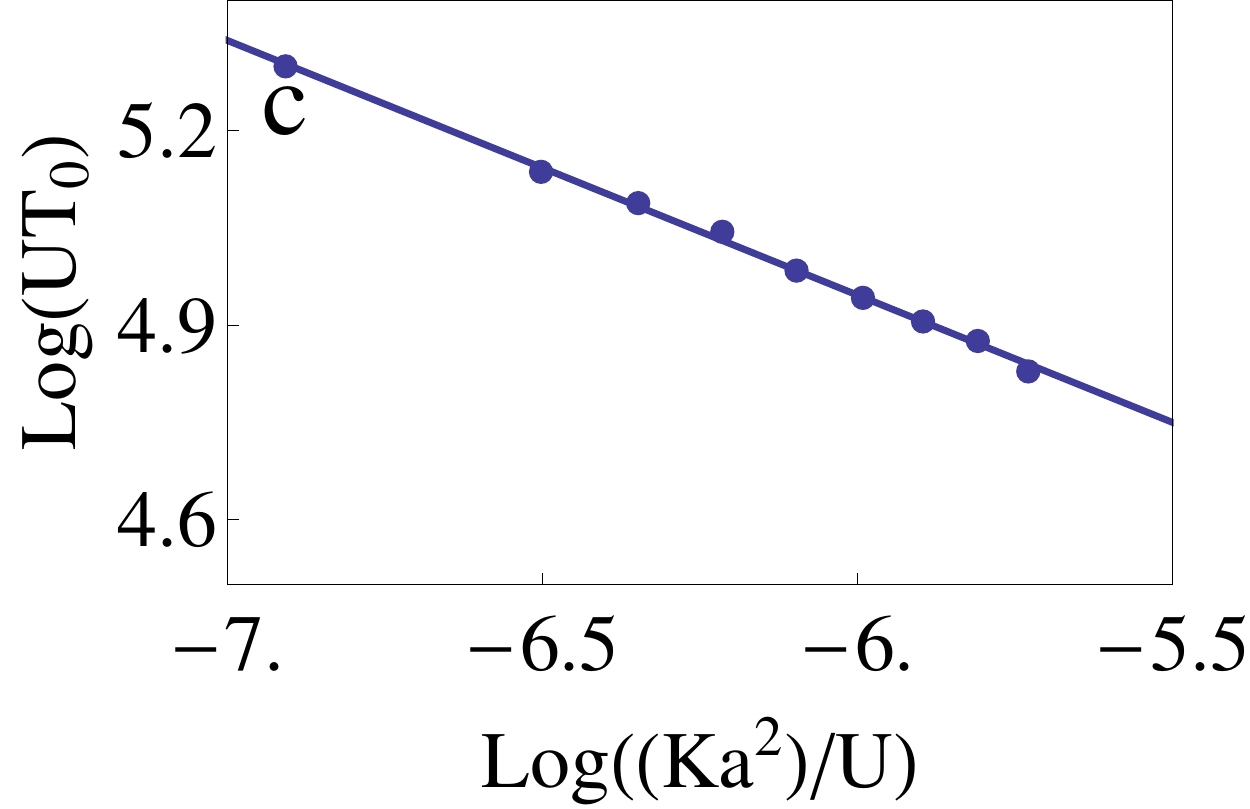}
\includegraphics[width=0.45\columnwidth]{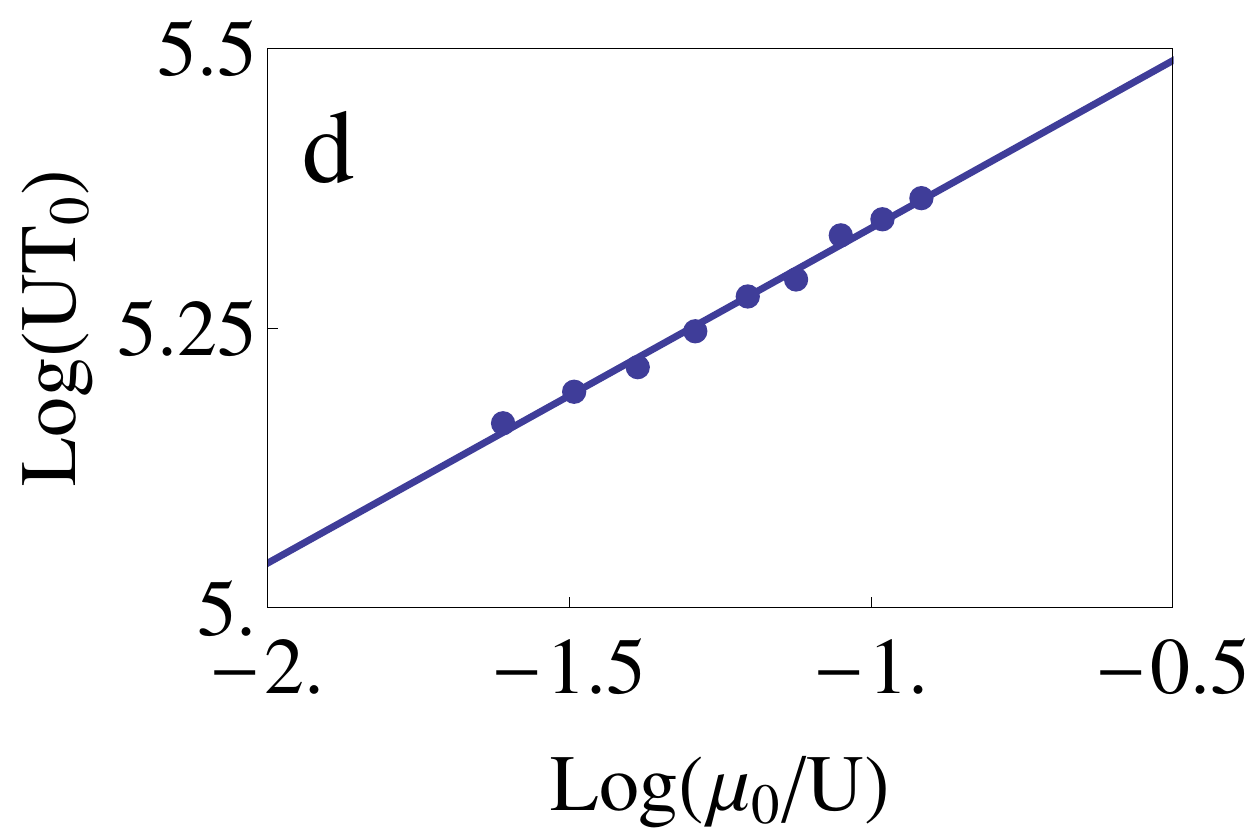}
\end{center}
\caption{(Color online) Time evolution of (a) $|\Delta_0(t)|$ and
(b) $\dot n \equiv \dot n_{r=0}(t)$ at the trap center after a
sudden quench of $J$ from $J_i=0.1 U$ to $J_f=0.02 U$. (c) and (d)
Plot of $\ln(T_0 U)$ as a function of $\ln(K a^2/U)$ and
$\ln(\mu_0/U)$ respectively. The straight lines in (c) and (d) show
linear fit to the data points.} \label{fig3}
\end{figure}

To obtain a qualitative understanding of this phenomenon, we note
that the trap leads to actual mass (particle) transport with a
typical velocity $v_b \simeq J_f/\hbar$ (where the optical lattice
spacing is set to unity) affecting the dynamics of the
 system\cite{collath1,natu1}. When a
boson hops outwards from a site $r$ along $(1,0)$ direction, it
encounters the boundary between the SF and the $n=0$ MI phase at
$\mu_{\bf r} = 0$ or at $r = {\rm Int}[\sqrt{2\mu_0/K}]$. Since the
latter phase is analogous to the boson vacuum, the bosons get
reflected back from this boundary. When the reflected wave reaches a
given point inside the boundary, it interferes with the coherent
oscillation and causes it to decohere. To see this process clearly,
we plot in Fig.~\ref{fig3}(a) and (b), the SF order parameter
$|\Delta|$ and the time derivative of the boson density, $\dot{n}$,
at the trap center as a function of time. We clearly see that the
sudden decoherence of $|\Delta(t)|$ coincides with an increase in
$\dot{n}$. The reflection of the bosons leads to inhomogeneous boson
flux at a given site leading, via continuity equation, to finite
$\dot n$ and hence to decoherence of the short time oscillation
pattern of $|\Delta_{r=0}(t)|$. For the central trap site, we expect
this to happen around $T_0 \simeq 2 r_x^{\rm max}/J_f = 2 {\rm
Int}[\sqrt{2\mu_0/K}]/J_f \simeq 200 U^{-1}$ \cite{comment0}. Our
reasoning above predicts $T_0 \sim 1/\sqrt{K}$ and $T_0 \sim
\sqrt{\mu_0}$; this is corroborated in Figs.\ \ref{fig3}(c) and
\ref{fig3}(d). These plots indicate $T_0 \sim K^{-0.53}$ and $T_0
\sim \mu^{0.41}$ which is roughly consistent with the behavior
$1/\sqrt{K}$ and $\sqrt{\mu_0}$ obtained from our qualitative
argument.

\begin{figure}[t!]
\begin{center}
\includegraphics[width=0.49\columnwidth]{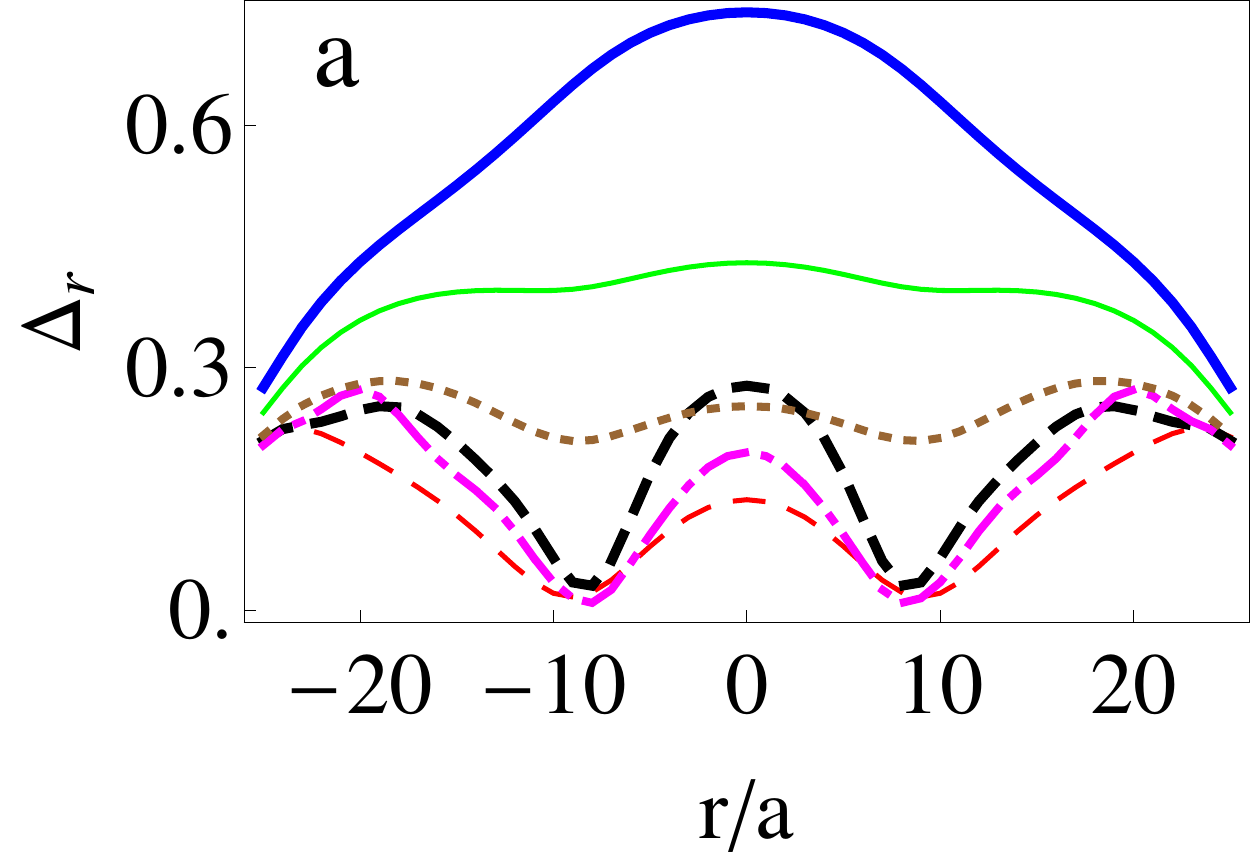}
\includegraphics[width=0.49\columnwidth]{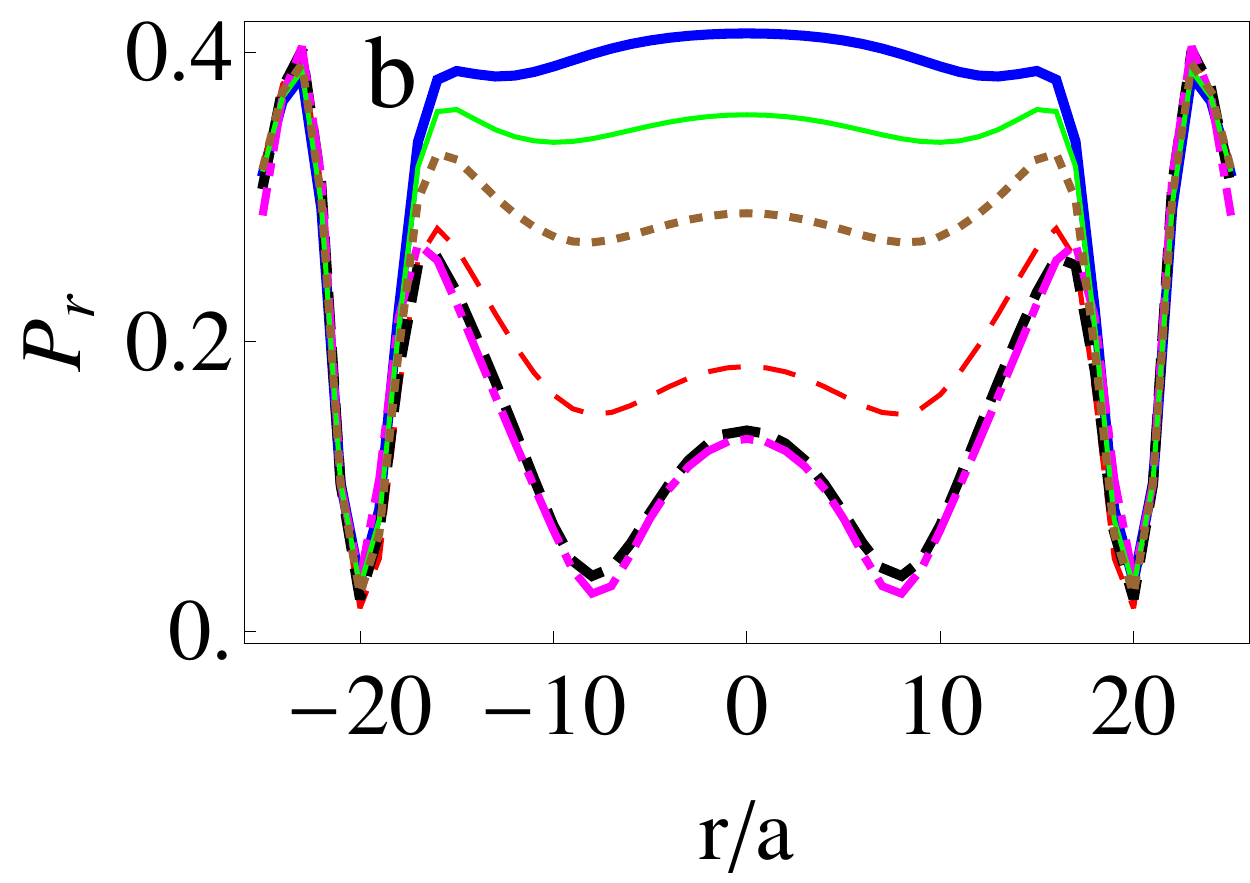}
\end{center}
\caption{(Color online) Spatial profile of (a)  the order parameter
amplitude $|\Delta_r|$ and (b) defect density $P_r$ after a linear
ramp of $J$ from $J_i=0.1 U$ to $J_f=0.02 U$. The different curves
correspond to $U\tau=1$ (blue solid line), $2$ (green solid line),
$3$ (brown dotted line), $5$ (orange dashed line), $10$ (pink
dash-dotted line), and $15$ (black dashed line). }\label{fig4}
\end{figure}

Finally, we study the effect of a linear ramp-down of the hopping
$J(t)=J_i +(J_f-J_i)t/\tau$ in this system. We choose $J_i(J_f)=0.1
(0.02)U$ so that the bosons start from the SF ground state and pass
through the equilibrium MI-SF transition for $r<14$. For $14\le r\le
20$ ($r>20$), the equilibrium state of the bosons are SF ($n=0$ MI)
state throughout the dynamics. We look at the system at the end of
the ramp ($t=\tau$) and study the local order parameter
$|\Delta_r(\tau)|$ and the local defect density $P_r(\tau)
=1-|\langle \psi_r(\tau)|\psi_{r G}\rangle|^2$, where
$|\psi_{rG}\rangle)$ is the ground state with $J=J_f$ and
$|\psi_{r}(\tau)\rangle)$ is the non-equilibrium state right after
the ramp. As shown in Fig.~\ref{fig4}(a) and (b), $|\Delta_r(\tau)|$
and $P_r(\tau)$ display non-monotonic spatial profiles for large
$\tau$ (close to adiabatic limit); they have a maximum at $r=0$
followed by an initial reduction and later enhancement as $r$
increases. We note that such spatial profiles have no analog for
trapped bosons in equilibrium.

To understand these novel profiles, we first consider the behavior
of $|\Delta_r(\tau)|$ and $P_r(\tau)$ near the trap center where the
effect of the trap potential is negligible. Here, for a slow quench,
one can use an adiabatic-impulse argument to estimate the deviation
of the final wavefunction from the initial one \cite{kz1}. The
evolution of the wavefunction gets the system in the adiabatic
regime (no defect production) when the instantaneous energy gap
$\epsilon(t)$ satisfies $d\epsilon/dt \le \epsilon(t)^2$. Near the
trap center, we can use LDA to define a local gap,
$\epsilon_r(t)={\rm Min}[E^p_r(t),E^h_r(t)]$, where $E^p_r(t)=
U-\mu_{\bf r} -2 z J(t)$ and $E^h_r(t)=\mu_{r} -zJ(t)$ are energies
of particle/hole production.
 The time at which the system enters the adiabatic regime, $t_1$, can be obtained
by solving $d\epsilon(t)/dt|_{t=t_1} =\epsilon^2(t_1)$ and yields
\begin{eqnarray}
t_1(r) &=& \tau [J_i-J_{cr}(r)](J_i-J_f)^{-1} +\sqrt{\tau/(J_i-J_f)}
\label{timp}
\end{eqnarray}
where $J_{cr}(r)={\rm Min}[(U-\mu_{\bf r})/2z,\mu_{r}/z]$ is the
equilibrium critical value of $J$ at the local effective potential
$\mu(r)$. For our choice of parameters, $J_{cr}(r)$ decreases with
$r$; consequently, $t_1$ is an increasing function of $r$. Thus the
trap center $r=0$ spends minimum time in the so called impulse
region, where the local wavefunction can adjust to the changes in
$J$. Thus the deviation of the wavefunction from the initial
superfluid state is minimum here. Hence
 $|\Delta_{r}(\tau)|$ has a maxima at $r=0$ and gradually decreases with $r$ with the profile of
$|\Delta_{r}(\tau)|$ having the same shape as $J_{cr}$. The defect
density profile near the trap center can also be understood from the
same argument if we remember that the defect density measures the
deviation of the state from the ground state with the final value of
$J$, and not from the initial state. Since the freeze-out occurs later
as we go outward, the states at larger $r$ are closer to the final
ground state and has less defect density.

However, the effect of larger time spent in the impulse region is
offset by the slower evolution due to the presence of the trap as we
move outward from the trap center. To understand this, consider the
Gutzwiller mean-field theory of trapped boson with $J=J(t)$. Using a
minimal model of three boson states near SI transition, $n=0,1,2$
per site \cite{supp1,mps1},
 the wavefunction  $\psi_{\rm mf} = \prod_{\bf r} \sum_{n_{\bf r}=0}^2 c_{n_{\bf r}}^{\bf
r}(t) |n\rangle$.
 In this limit, $\Delta_{\bf r}(t) = \Delta_{1{\bf
r}}(t) e^{-i \mu_{\bf r} t/\hbar} +\Delta_{2{\bf r}}(t)
e^{-i(U-\mu_{\bf r})t/\hbar}$\cite{supp1}, where, within a rotating
wave approximation,
\begin{eqnarray}
\dot \Delta_{1[2]{\bf r}}(t) &=& -i J(t) \sum_{\langle {\bf
r'}\rangle} |c_{1}^{\bf r'}|^2 A_{{\bf r r'}}(t) [B_{{\bf r r'}}(t)]
e^{i \delta
\mu_{\ell}t/\hbar}
\label{mftorder}
\end{eqnarray}
Here $\ell$ denotes the link between ${\bf r}$ and ${\bf r'}$, and
the coefficients $A_{\bf r r'}= c_0^{{\bf r'}\ast} c_1^{\bf r}$ and
$B_{\bf r r'}= \sqrt{2} c_1^{{\bf r}\ast} c_2^{\bf r'}$. It can be
seen from Eq.\ \ref{mftorder} that for $t>t_0$, where $t_0$ is
defined by $J(t_0)= \delta \mu_{\ell}$, the oscillatory terms would
wash out the dynamics of $|\Delta|$. Thus  one expects the evolution
of $\Delta_{r}$ to be small at sites for which $t_0$ is small
compared to the ramp time $\tau$,. It is easy to see numerically
that $J(t_0) = J_f$ for $r\simeq 9$. It is approximately around this
point that the slow nature of the evolution overcompensates for
larger time spent in the impulse region and one finds an upturn of
the order parameter. Further, we expect the dynamics to become
progressively slower as we move further outwards. Consequently,
$|\Delta_{r}(\tau)|$ increases with $r$ for $8\le r\le 19$. A
similar behavior is seen for the defect density $P_r$. For $r>19$,
the small initial $|\Delta_r(0)|$, owing to the proximity of the
system to the $n=0$ MI phase, outweighs the slow evolution during
ramp dynamics and thus both $|\Delta_r(\tau)|$ and $P_r(\tau)$ again
starts to decrease.

Experimental verification of our theory can be most easily done by
using standard parity of occupation measurements \cite{bakr1,veng1}
which would capture the non-monotonic dependence of $P_r$ after the
ramp. This would require inference of $P_r$ from the experimentally
measured parity of boson occupation. We note that this can be easily
done using techniques of Ref.\ \cite{bakr1} since only states with
$n_r\le 2$ bosons have appreciable weights in the regime $U \gg J$.

In conclusion, we have presented a novel hopping expansion
techniques which allows us to address the non-equilibrium dynamics
of trapped bosons in the strong coupling regime beyond mean-field
theory. We have identified a trap-induced length scale $r_{max}\sim
1/\sqrt{K}$, which acts as a reflection boundary for the dynamics of
the bosons after the quench leading to a sudden decoherence of the
collapse revival oscillations at $T_0 \sim K^{-1/2}$. In evolution
under a slow linear ramp of $J$, we have found a novel non-monotonic
spatial profile of the defect density and order parameter amplitude
which is qualitatively different from its counterpart in equilibrium
and can be measured easily by available experiments.

AD acknowledges helpful discussions with Juan Carrasquilla at early
stages of this project.

\section{Supplementary Material:Sketch of the variational energy calculation}

In this section, we provide a sketch of the calculation of the
variational energy $E_G = \langle \psi'|H_{\rm eff} |\psi'\rangle$
used in the main text. We first write the effective Hamiltonian as
\begin{eqnarray}
H_{\rm eff} &=& H_{\rm eff}^0 + H_{\rm eff}^1 + H_{\rm eff}^2,
\nonumber\\
H_{\rm eff}^0 &=& \sum_{\bf r} \left[-\mu_{\bf r} {\hat n}_{\bf r} +
\frac{U}{2} {\hat
n}_{\bf r}({\hat n}_{\bf r} -1)\right] \nonumber\\
H_{\rm eff}^1 &=& \sum_{\ell, i=1,2} T_{\ell i}^{{\bar \alpha}_{\ell
i}}
\eta_{{\bar \alpha}_{\ell i}} \nonumber\\
H_{\rm eff}^2  &=&  \sum_{\ell \ell'} \sum_{\alpha \ne {\bar
\alpha}_{\ell i}} \sum_{i,j=1,2} \eta_{{\bar \alpha}_{\ell i}}
\left[ T_{\ell'j}^{\alpha}, T_{\ell i}^{{\bar
\alpha}_{\ell i}}\right]/\Delta E_{\ell'j}^{\alpha}  \label{app1a}\\
&& + \sum_{\ell \ell'} \sum_{\alpha \ne {\bar \alpha}_{\ell i},
\alpha' \ne {\bar \alpha}_{\ell' j}} \sum_{i,j=1,2} \left[ T_{\ell
i}^{\alpha}, T_{\ell' j}^{\alpha'} \right]/(2\Delta E_{\ell
i}^{\alpha}), \nonumber\
\end{eqnarray}
where we have introduced the link variable $\ell$ between two
neighboring sites ${\bf r}$ and ${\bf r'}$, $\ell'$ denote either
the same link as or the nearest neighbor link of $\ell$, $\Delta
E_{\ell}^{\alpha}= \alpha U-\mu_{\bf r}+\mu_{\bf r'}$, and
$\eta_{{\bar \alpha}_{\ell i}}=0(1)$ if ${\bar \alpha}_{\ell i}$
exists (does not exist) for a given link. The definition of all
other operators used in Eq.\ \ref{app1a} is given in the main text.
In terms of these link variables, one can distinguish between the
outward and inward hopping processes on link as $T_{\ell 1}$ and
$T_{\ell 2}$ respectively. The hopping term can be then written as
$T = \sum_{\ell} T_{\ell} = \sum_{\ell \alpha} T_{\ell 1}^{\alpha} +
T_{\ell 2}^{\alpha}$, where
\begin{eqnarray}
T_{\ell 1}^{\alpha} &=& -J \sum_{n_{\bf r}} \sqrt{n_{\bf r}(n_{\bf
r} + \alpha)} |n_{\bf r}-1, n_{\bf r} + \alpha \rangle
\nonumber\\
&& \times \langle
n_{\bf r},n_{\bf r} + \alpha-1|, \nonumber\\
T_{\ell 2}^{\alpha} &=& -J \sum_{n_{\bf r}}\sqrt{(n_{\bf
r}+1)(n_{\bf r} +1- \alpha)} |n_{\bf r}+1, n_{\bf r} - \alpha
\rangle \nonumber\\
&& \times \langle n_{\bf r},n_{\bf r} - \alpha +1|.  \label{hopp1}
\end{eqnarray}
We note that Eq. \ref{hopp1} is analogous to Eq.\ (2) of the main
text. Also, it is worth noting that in this representation, one can
write
\begin{eqnarray}
iS &=& \sum_{\ell} \sum_{i=1,2} \sum_{\alpha \ne {\bar \alpha}_{\ell
i}}T_{\ell i}^{\alpha}/\Delta E_{\ell i}^{\alpha}. \label{candef}
\end{eqnarray}
which has been used to derive Eq.\ \ref{app1a}. Using Eq.\
\ref{app1a}, one can express the variational ground state energy as
\begin{eqnarray}
E_G = \sum_{\alpha =0,1,2} E_G^{\alpha}, \quad E_G^{\alpha} =
\langle \psi'|H_{\rm eff}^{\alpha} |\psi'\rangle, \label{app1b}
\end{eqnarray}
where $|\psi'\rangle$ is the Gutzwiller wavefunction given by
\begin{eqnarray}
|\psi'\rangle = \prod_{\bf r} \sum_{n_{\bf r}} f_{n_{\bf r}}^{\bf r}
|n_{\bf r}\rangle \label{app1c}
\end{eqnarray}
We begin with computation $E_G^0$. Since this term involves only
density operators on a single site, it is straightforward to see
that
\begin{eqnarray}
E_G^0 &=& \sum_{n_{\bf r}} \sum_{{\bf r}} |f_{n_{\bf r}}^{\bf r}|^2
\left[ -\mu_{\bf r} n_{\bf r} + \frac{U}{2} n_{\bf r}(n_{\bf r}-1)
\right] \label{app1d}
\end{eqnarray}

To compute $E_G^{1}$, which involves terms involving a single link
and hence two adjacent sites, we first note that using Eq. 2 of the
main text, one can write
\begin{eqnarray}
T_{\ell 1}^{\alpha} | n_1, n_2 \rangle &=& -J \sum_{n_{\bf r}}
\delta_{n_{\bf r}, n_1} \delta_{n_{\bf r} +\alpha-1,n_2}
\sqrt{n_{\bf r}(n_{\bf r}+\alpha)} \nonumber\\
&& \times |n_{\bf r}-1, n_{\bf
r}+\alpha \rangle \nonumber\\
T_{\ell 2}^{\alpha} | n_1, n_2 \rangle &=& -J \sum_{n_{\bf r}}
\delta_{n_{\bf r}, n_1} \delta_{n_{\bf r} - \alpha+1,n_2}
 \nonumber\\
&& \times \sqrt{(n_{\bf r}+1)(n_{\bf r}-\alpha)} |n_{\bf r}+1,
n_{\bf r}-\alpha \rangle \label{app1e}
\end{eqnarray}
where ${\bf r}$ denotes a site of the link $\ell$. Using Eq.\
\ref{app1e}, one can obtain $E_G^1 = \langle \psi' \sum_{\ell,
i=1,2} T_{\ell i}^{{\bar \alpha}_{\ell i}} \eta_{{\bar \alpha}_{\ell
i}} |\psi' \rangle$ to be
\begin{widetext}
\begin{eqnarray}
E_G^{1} &=& -J \sum_{n_{\bf r} \ell} \left[ \eta_{{\bar
\alpha}_{\ell 1}} \sqrt{n_{\bf r} (n_{\bf r} + {\bar \alpha}_{\ell
1} -1)} f_{n_{\bf r}-1}^{\ast \bf r} f_{n_{\bf r}}^{\bf r} f_{n_{\bf
r}+{\bar \alpha}_{\ell 1}}^{\ast \bf r'} f_{n_{\bf r}+{\bar
\alpha}_{\ell 1}-1}^{\bf r'} + \eta_{-{\bar \alpha}_{\ell
1}}\sqrt{n_{\bf r} (n_{\bf r} + {\bar \alpha}_{\ell 1} + 1)}
f_{n_{\bf r}+1}^{\ast \bf r} f_{n_{\bf r}}^{\bf r} f_{n_{\bf
r}+{\bar \alpha}_{\ell 1}}^{\ast \bf r'} f_{n_{\bf r}+{\bar
\alpha}_{\ell 1}+1}^{\bf r'} \right], \nonumber\\ \label{app1f}
\end{eqnarray}
\end{widetext}
where we have used ${\bar \alpha}_{\ell 2}= -{\bar \alpha}_{\ell
1}$, ${\bf r'}$ is the neighboring site of ${\bf r}$, and $\ell$
denotes the link between these two sites .

Next, we discuss computation of terms $O(J^2/U^2)$. First, we
consider the computation of terms in $H_2$ for which $\ell=\ell'$.
For computation of these terms, one uses the identities
\begin{widetext}
\begin{eqnarray}
T_{\ell 1}^{\alpha} T_{\ell 1}^{\beta} |n_1,n_2\rangle &=& J^2
\delta_{\beta, \alpha-2} \delta_{n_2,n_1+\alpha-3}
\sqrt{n_1(n_1-1)(n_2+1)(n_2+2)} |n_1-2,n_2+2\rangle \nonumber\\
T_{\ell 1}^{\alpha} T_{\ell 2}^{\beta} |n_1,n_2\rangle &=& J^2
\delta_{\beta, -\alpha} \delta_{n_2,n_1+\alpha+1}
n_2(n_1+1) |n_1,n_2\rangle \nonumber\\
T_{\ell 2}^{\alpha} T_{\ell 1}^{\beta} |n_1,n_2\rangle &=& J^2
\delta_{\beta, -\alpha} \delta_{n_2,n_1-\alpha-1}
n_1 (n_2+1) |n_1,n_2\rangle \nonumber\\
T_{\ell 2}^{\alpha} T_{\ell 2}^{\beta} |n_1,n_2\rangle &=& J^2
\delta_{\beta, \alpha-2} \delta_{n_2,n_1-\alpha+3}
\sqrt{(n_1+1)(n_1+2)(n_2-1)n_2} |n_1+2,n_2-2\rangle. \label{app1g}
\end{eqnarray}
\end{widetext}
For example, using these one can compute one of the representative
terms in the expression of $E_G^2$ as $E_{G(1)}^2 = \langle \psi'|
\sum_{\ell \alpha \ne \pm {\bar \alpha}_{\ell 1}} T_{\ell
1}^{\alpha} ( T_{\ell 1}^{{\bar \alpha}_{\ell 1}} + T_{\ell
2}^{-{\bar \alpha}_{\ell 1}})/\Delta E_{\ell
1}^{\alpha}|\psi'\rangle$. A few lines of straightforward algebra
yields
\begin{widetext}
\begin{eqnarray}
E_{G(1)}^2 &=& \sum_{\ell, n_{\bf r} }\sum_{\alpha \ne \pm {\bar
\alpha}_{\ell 1}}  \frac{J^2 \eta_{{\bar \alpha}_{\ell 1}}}{\Delta
E_{\ell_1}^{{\bar \alpha}_{\ell 1} +2}} f_{n_{\bf r}-2}^{\ast \bf r}
f_{n_{\bf r}}^{\bf r} f_{n_{\bf r}+{\bar \alpha}_{\ell 1}}^{\ast \bf
r'} f_{n_{\bf r}+{\bar \alpha}_{\ell 1}-1}^{\bf r'}  \sqrt{n_{\bf r}
(n_{\bf r}+1) (n_{\bf r} + {\bar \alpha}_{\ell 1}) (n_{\bf r} +
{\bar \alpha}_{\ell 1} +1)} \label{app1h}
\end{eqnarray}
\end{widetext}
All others terms with $\ell=\ell'$ which contribute to $E_G^2$ can
be computed in an analogous fashion.

Next, we consider the terms in $E_G^{2}$ which originates from terms
in $H_2$ which has $\ell$ and $\ell'$ as neighboring links. For
these terms, one involves three lattice sites which have coordinates
${\bf r}$, ${\bf r'}$, and ${\bf r}"$. We use the convention that
the link between sites with coordinates ${\bf r}$ and ${\bf r'}$ is
$\ell$ and that ${\bf r'}$ is the middle site connecting the link
$\ell$ and $\ell'$. With this convention, there are two claases of
hopping terms. In the first of these classes, $T_{\ell'}$ acts on
the state $|\psi'\rangle$ before $T_{\ell}$ leading to the following
relations:
\begin{widetext}
\begin{eqnarray}
T_{\ell 1}^{\alpha} T_{\ell' 1}^{\beta}|n_1,n_2,n_3\rangle  &=& J^2
\delta_{n_2, n_1+\alpha} \delta_{n_3,n_2 +\beta -1} n_2
\sqrt{n_1(n_3+1)} |n_1-1,n_2,n_3+1\rangle \nonumber\\
T_{\ell 1}^{\alpha} T_{\ell' 2}^{\beta}|n_1,n_2,n_3\rangle  &=& J^2
\delta_{n_2, n_1+\alpha-2} \delta_{n_3,n_2 -\beta +1}
\sqrt{(n_2+1)(n_2+2)n_1 n_3} |n_1-1,n_2+2,n_3-1\rangle \nonumber\\
T_{\ell 2}^{\alpha} T_{\ell' 1}^{\beta}|n_1,n_2,n_3\rangle  &=& J^2
\delta_{n_2, n_1-\alpha+2} \delta_{n_3,n_2 -\beta +1}
\sqrt{n_2(n_2-1)(n_1+1)(n_3+1)} |n_1+1,n_2-2,n_3+1\rangle \nonumber\\
T_{\ell 2}^{\alpha} T_{\ell' 2}^{\beta}|n_1,n_2,n_3\rangle  &=& J^2
\delta_{n_2, n_1-\alpha} \delta_{n_3,n_2 -\beta +1} n_2
\sqrt{n_3(n_1+1)} |n_1+1,n_2,n_3-1\rangle. \label{app1i}
\end{eqnarray}
\end{widetext}
In the second class of terms, $T_{\ell}$ acts on $|\psi'\rangle$
before $T_{\ell'}$ and this leads to
\begin{eqnarray}
T_{\ell' 1}^{\alpha} T_{\ell 1}^{\beta}|n_1,n_2,n_3\rangle  &=&
T_{\ell 1}^{\beta -1} T_{\ell' 1}^{\alpha+1}|n_1,n_2,n_3\rangle,  \nonumber\\
T_{\ell' 1}^{\alpha} T_{\ell 2}^{\beta}|n_1,n_2,n_3\rangle  &=&
T_{\ell 2}^{\beta + 1} T_{\ell' 1}^{\alpha-1}|n_1,n_2,n_3\rangle, \nonumber\\
T_{\ell' 2}^{\alpha} T_{\ell 1}^{\beta}|n_1,n_2,n_3\rangle  &=&
T_{\ell 1}^{\beta +1} T_{\ell' 2}^{\alpha-1}|n_1,n_2,n_3\rangle, \nonumber\\
T_{\ell' 2}^{\alpha} T_{\ell 2}^{\beta}|n_1,n_2,n_3\rangle  &=&
T_{\ell 2}^{\beta -1} T_{\ell' 2}^{\alpha+1}|n_1,n_2,n_3\rangle.
\label{app1j}
\nonumber\\
\end{eqnarray}
Using these identities one can evaluate the contribution of all
terms in $H_2$ with $\ell \ne \ell'$ to $E_G^2$ . A representative
example of such a term is $E_{G(2)}^2 = \langle \psi'| \sum_{\langle
\ell \ell'\rangle} \sum_{\alpha \ne \pm {\bar \alpha}_{\ell 1}}
T_{\ell 1}^{\alpha} ( T_{\ell' 1}^{{\bar \alpha}_{\ell' 1}} +
T_{\ell' 2}^{-{\bar \alpha}_{\ell' 1}})/\Delta E_{\ell
1}^{\alpha}|\psi'\rangle$. A straightforward calculation using Eq.\
\ref{app1i} yields
\begin{widetext}
\begin{eqnarray}
E_{G(2)}^2 &=& \sum_{\langle \ell \ell'\rangle, n_{\bf r}}
\sum_{\alpha \ne {\bar \alpha}_{\ell' 1}} \frac{J^2}{\Delta E_{\ell
1}^\alpha} \left[ \eta_{{\bar \alpha}_{\ell' 1}} f_{n_{\bf
r}-1}^{\ast \bf r} f_{n_{\bf r}}^{\bf r} |f_{n_{\bf r}
+\alpha}^{{\bf r'}}|^2 f_{n_{\bf r}+\alpha+ {\bar \alpha}_{\ell'
1}}^{\ast \bf r"} f_{n_{\bf r}+\alpha + {\bar \alpha}_{\ell'
1}-1}^{\bf r"} (n_{\bf r} + \alpha) \sqrt{n_{\bf r} (n_{\bf r} +
\alpha +{\bar \alpha}_{\ell' 1})} \right. \\
&& \left. + \eta_{-{\bar \alpha}_{\ell' 1}} f_{n_{\bf r}-1}^{\ast
\bf r} f_{n_{\bf r}}^{\bf r} f_{n_{\bf r} +\alpha}^{\ast {\bf r'}}
f_{n_{\bf r}+\alpha-1}^{\bf r'}  f_{n_{\bf r}+\alpha+ {\bar
\alpha}_{\ell' 1}-2}^{\ast \bf r"} f_{n_{\bf r}+\alpha + {\bar
\alpha}_{\ell' 1}-1}^{\bf r"} \sqrt{n_{\bf r} (n_{\bf r} +
\alpha)(n_{\bf r} + \alpha -1) (n_{\bf r} + \alpha +{\bar
\alpha}_{\ell' 1}-1)}\right] \label{app1k} \nonumber
\end{eqnarray}
\end{widetext}
All other terms with $\ell\ne \ell'$ can be obtained in a similar
fashion using Eqs.\ \ref{app1i} and \ref{app1j}. Together, these
terms leads to the expression of $E_G^2$ used in the main text. The
ground state wavefunction, used for obtaining Fig. 1 in the main
text, is obtained by numerical minimization of $E_G$.

To obtain the numerical solution of the time-depedent Schrdinger
equation $(i \hbar \partial_t +\partial_t S[J(t)])|\psi'\rangle=
H_{\rm eff}[J(t)] |\psi'\rangle$ used in the main text, we first
write $|\psi'\rangle = \prod_{\bf r} \sum_{n_{\bf r}} f^{\bf
r}_{n_{\bf r}}(t) |n_{\bf r}\rangle$. Using standard procedure
\cite{stc2}, it is then straightforward to obtain the equations for
time evolution of $f^{{\bf r}}_{n_{\bf r}}(t)$ which is given by
\begin{eqnarray}
&& i\hbar \partial_t f_{n_{\bf r}}^{\bf r}(t) = \delta
E_G[\{f_{n_{\bf r}}^{\bf r}\}; J(t)]/\delta f_{n_{\bf r}}^{\ast \bf
r}(t) + i \hbar
\dot J(t) \nonumber\\
&& \times \sum_{\alpha \langle {\bf r'}\rangle} \left[\sqrt{n_{\bf
r}} f_{n_{\bf r}-1}^{\bf r} \phi_{n_{\bf r}-\alpha}^{{\bf r'}} +
\sqrt{n_{\bf r}+1} f_{n_{\bf r}+1}^{{\bf r}} \phi_{n_{\bf
r}+\alpha}^{{\bf r'}\ast} \right] \nonumber\\
&& \times \sum_{i=1,2} (1-\delta_{\alpha {\bar \alpha}_{\ell
i}^{\alpha}})/ \Delta E_{\ell i}^{\alpha}. \label{scheq1}
\end{eqnarray}
From the expression of $E_G$ obtained earlier, one can obtain, after
some straightforward, but tedious algebra the terms $\delta E_G
/\delta f_{n_{\bf r}}^{\ast {\bf r}}$. Eq.\ \ref{scheq1} is then
numerically solved to obtain $f^{{\bf r}}_{n_{\bf r}}(t)$. This
leads to $|\psi'(t)\rangle$, and consequently to $|\Delta_{\bf
r}(t)|$ and $\delta n_{\bf r}(t)$ used in the main text.

\section{Supplementary Materials: Mean-field theory}

In this section, we present the mean-field calculation leading to
Eq.\ 8 of the main text. To do this, let us consider the mean-field
Bose-Hubbard Hamiltonian given by
\begin{eqnarray}
H_{\rm mf} &=& H_0 + \sum_{\bf r} ( \Delta_{\bf r}(t) b_{\bf
r}^{\dagger} + {\rm h.c.}) \nonumber\\
\Delta_{\bf r} &=& -J(t) \sum_{\langle {\bf r'} \rangle} \langle
b_{\bf r'} \rangle \label{app2a}
\end{eqnarray}
where $\langle {\bf r'}\rangle$ denotes nearest neighbor sites ${\bf
r'}$ to ${\bf r}$, the time dependence of $J(t)$ is kept arbitrary
for now, and the expectation is taken using the time-dependent
variational Gutzwiller wavefunction $|\psi\rangle = \prod_{\bf r}
\sum_{n_{\bf r}} c_{n_{\bf r}}^{\bf r}(t) |n_{\bf r}\rangle$. The
corresponding mean-field equations for $c_{n_{\bf r}}^{\bf r}(t)$ is
given by
\begin{eqnarray}
&& (i \partial_t - \epsilon_{n_{\bf r}}^{\bf r}) c_{n_{\bf r}}^{\bf
r} = -J(t) \sum_{\langle {\bf r'} \rangle n_{\bf r'}} \left[ \sqrt{
n_{\bf r} (n_{\bf r'}+1)} c_{n_{\bf r'}}^{\ast \bf r'} c_{n_{\bf
r'}+1}^{\bf r'} \right. \nonumber\\ && \left. \times c_{n_{\bf
r}-1}^{\bf r}+ \sqrt{ n_{\bf r'} (n_{\bf r}+1)} c_{n_{\bf r'}}^{\ast
\bf r'} c_{n_{\bf r'}-1}^{\bf r'} c_{n_{\bf r}+1}^{\bf r}
\right],\label{app2b}
\end{eqnarray}
where $\epsilon_{n_{\bf r}}^{\bf r}= -\mu_{\bf r} n_{\bf r} + U
n_{\bf r}(n_{\bf r}-1)/2$ is the local on-site energy.

To proceed further, we note that for $zJ(t)/U \ll 1$ (where $z$ is
the coordination number o the lattice) and for $0<\mu_{\bf r}/U <1$
(which holds for all $|{\bf r}| \le 20$ in the present case), one
has $ c_{n_{\bf r}>2}^{\bf r}=0$ and $|c_{1}^{\bf r}| \gg |c_0^{\bf
r}|, |c_{2}^{\bf r}|$. One can then approximate the mean-field
equations for $c_{n_{\bf r} \le 2}^{\bf r}$ to be
\begin{eqnarray}
i \partial_t c_{0}^{\bf r} &=& -J(t) \sum_{\langle {\bf r'} \rangle}
c_1^{\bf r} (c_{1}^{\ast \bf r'} c_{0}^{\bf r'} + \sqrt{2}
c_{2}^{\ast \bf r'} c_{1}^{\bf r'} )
\nonumber\\
(i \partial_t +\mu_{\bf r}) c_{1}^{\bf r} &=& -J(t) \sum_{\langle
{\bf r'}\rangle} \left[c_0^{\bf r} ( \sqrt{2} c_{1}^{\ast \bf r'}
c_{2}^{\bf r'}
+ c_0^{\ast \bf r'} c_1^{\bf r'}) \right. \nonumber\\
&& \left. c_{2}^{\bf r} (\sqrt{2} c_{1}^{\ast \bf r'} c_{0}^{\bf r'}
+ 2 c_2^{\ast \bf r'} c_{1}^{\bf r'}) \right], \\
(i \partial_t - U +\mu_{\bf r}) c_{2}^{\bf r} &=& -J(t)
\sum_{\langle {\bf r'} \rangle} c_1^{\bf r} (2 c_{2}^{\ast \bf r'}
c_{1}^{\bf r'} + \sqrt{2} c_{1}^{\ast \bf r'} c_{0}^{\bf r'} ),
\label{app2c} \nonumber\
\end{eqnarray}
We now define the slow variables ${\tilde c}_{n_{\bf r}}^{\bf r} =
c_{n_{\bf r}} \exp[-i(\epsilon_{n_{\bf r}}^{\bf r} +\mu_{\bf r}) t]$
and obtain their equation of motion as
\begin{eqnarray}
i \partial_t {\tilde c}_{0}^{\bf r} &=& -J(t) \sum_{\langle {\bf r'}
\rangle} {\tilde c}_1^{\bf r} e^{- i \delta \mu_{{\bf r r'}} t}
\left( {\tilde c}_{1}^{\ast \bf r'} {\tilde c}_{0}^{\bf r'}  +
\sqrt{2} {\tilde c}_{2}^{\ast \bf r'} {\tilde c}_{1}^{\bf r'} e^{i U
t} \right)
\nonumber\\
i \partial_t {\tilde c}_{1}^{\bf r} &=& -J(t) \sum_{\langle {\bf
r'}\rangle} \left[{\tilde c}_0^{\bf r} e^{i \delta \mu_{{\bf r r'}}
t} \left(\sqrt{2} {\tilde c}_{1}^{\ast \bf r'} {\tilde c}_{2}^{\bf
r'} e^{-iUt}
+ {\tilde c}_0^{\ast \bf r'} {\tilde c}_1^{\bf r'}\right) \right.\nonumber\\
&& \left. + {\tilde c}_{2}^{\bf r} e^{-i \delta \mu_{{\bf r r'}}
t}\left(\sqrt{2} {\tilde c}_{1}^{\ast \bf r'} {\tilde c}_{0}^{\bf
r'}
e^{-iUt} + 2 {\tilde c}_2^{\ast \bf r'} {\tilde c}_{1}^{\bf r'}\right)\right], \\
i \partial_t {\tilde c}_{2}^{\bf r} &=& -J(t) \sum_{\langle {\bf r'}
\rangle} {\tilde c}_1^{\bf r} e^{i \delta \mu_{{\bf r r'}}t}(2
{\tilde c}_{2}^{\ast \bf r'} {\tilde c}_{1}^{\bf r'} + \sqrt{2}
{\tilde c}_{1}^{\ast \bf r'} {\tilde c}_{0}^{\bf r'} e^{iUt} ),
\label{app2d} \nonumber\
\end{eqnarray}
where $\delta \mu_{{\bf r r'}} \equiv \delta \mu_{\ell} = \mu_{\bf
r'}-\mu_{\bf r}$.

Next, we compute the expectation value of the superfluid order
parameter: $\Delta_{\bf r} = \langle \psi|b_{\bf r}|\psi\rangle$.
Using the expression for $|\psi\rangle$, one obtains
\begin{eqnarray}
\Delta_{\bf r} &=& {\tilde c}_0^{\ast {\bf r}} {\tilde c}_1^{\bf r}
e^{-i \mu_{\bf r} t} + \sqrt{2} {\tilde c}_1^{\ast {\bf r}} {\tilde
c}_2^{\bf r} e^{-i (U-\mu_{\bf r})t} \nonumber\\
&=& \Delta_{1{\bf r}} e^{-i \mu_{\bf r} t} + \Delta_{2{\bf r}} e^{-i
(U-\mu_{\bf r}) t} \label{app2e}
\end{eqnarray}
Using the expressions of $\Delta_{1(2) {\bf r}}$ in Eq.\
\ref{app2e}, one can now obtain there equations of motion from those
of ${\tilde c}_{n}^{\bf r}$ ($n=0,1,2$) obtained in Eq.\
\ref{app2d}. A few lines of straightforward algebra shows that in
the limit when $|{\tilde c}_{1}^{\bf r}|^2 \gg |{\tilde c}_{0}^{\bf
r}|^2, |{\tilde c}_2^{\bf r}|^2$, one has
\begin{eqnarray}
i \partial_t \Delta_{1{\bf r}} &=& J(t) \sum_{\langle {\bf
r'}\rangle} |{\tilde c}_1^{\bf r}|^2 \left( A_{\bf r r'} + e^{-iU t}
B_{\bf r r'} \right) e^{-i \delta \mu_{\bf r r'} t} \nonumber\\
i \partial_t \Delta_{2{\bf r}} &=& -\sqrt{2} J(t) \sum_{\langle {\bf
r'}\rangle} |{\tilde c}_1^{\bf r}|^2 \left( B_{\bf r r'} + e^{iU t}
A_{\bf r r'} \right) e^{-i \delta \mu_{\bf r r'} t} \nonumber\\
\label{app2f}
\end{eqnarray}
where $A_{\bf r r'} = {\tilde c}_0^{\ast \bf r'} {\tilde c}_1^{\bf
r}$ and $B_{\bf r r'} = \sqrt{2} {\tilde c}_1^{\ast \bf r'} {\tilde
c}_2^{\bf r}$. Using the rotating wave approximation, one can drop
the terms with the factor $e^{iUt/\hbar}$ in right side of Eq.\
\ref{app2f}. This yields Eq.\ 8 in the main text where we have
reexpressed $A_{\bf r r'}$ and $B_{\bf r r'}$ in terms of $c_n^{\bf
r}$.

\vspace{-\baselineskip}


\begin{thebibliography}{99}
\bibitem{footnotegamma} We use a value of $\gamma=2$; we have
checked that our results are insensitive to the exact choice of
$\gamma$.
\bibitem{commentzz} This procedure is justified as long
as $zJ(t)/U \ll1$ for all $t$ and thus can be used to study the
dynamics of the bosons in the SF region near the critical point.
\end{thebibliography}
\end{document}